\documentclass[twocolumn,showpacs,prd,aps,rsi,floatfix,superscriptaddress,showkeys,citeautoscript,longbibliography]{revtex4-2}
\usepackage{hyperref}
\usepackage{graphicx,wrapfig,epstopdf,booktabs,soul,array,lmodern}
\usepackage[left=1.65cm,right=1.65cm,top=1.95cm,bottom=1.95cm]{geometry}
\usepackage{balance}
\usepackage{epigraph}
\usepackage{footmisc}
\usepackage{dcolumn}
\usepackage{dutchcal}
\usepackage{bm,color}
\usepackage{amsmath,amssymb,dsfont,amstext,amsfonts}
\usepackage{extarrows}
\usepackage{changepage}
\usepackage{lettrine}
\usepackage{times,mathptmx}
\usepackage[english]{babel}
\usepackage{mathtools}
\usepackage{bbold}
\usepackage{wasysym}
\usepackage{mathtools}
\usepackage{bbold} 
\usepackage{braket}
\usepackage{dsfont}
\usepackage{bigints}
\usepackage{soul}
\usepackage{xcolor}
\usepackage{amssymb,amsmath,amsthm,amssymb,multirow,printlen,layouts}
\usepackage{color}
\definecolor{lightgray}{gray}{0.5}
\usepackage[export]{adjustbox}
\usepackage{footnote}
\usepackage{ucs}
\usepackage{braket}
\usepackage{dsfont}
\usepackage{bigints}

\usepackage{soul}
\usepackage{url,hyperref}
\usepackage[font=footnotesize,labelfont=bf]{caption}
\usepackage{charter}
\usepackage[T1]{fontenc}
\usepackage[usenames,dvipsnames,svgnames,table]{xcolor}
\definecolor{CustomBlue}{rgb}{0.2, 0.18, 0.65} 
\definecolor{CustomRed}{rgb}{0.8, 0.2, 0.2} 
\usepackage{xcolor}
\definecolor{mydarkred}{rgb}{0.7, 0.0, 0.0}  
\hypersetup{
    backref=true,
    pagebackref=true,
    colorlinks=true,
    linkcolor=CustomBlue,
    citecolor=CustomBlue,
    urlcolor=CustomBlue
}

\usepackage{float}
\usepackage{fancyhdr}
\usepackage{orcidlink}
\usepackage{fnpos}
\addto{\captionsenglish}{%
  
}
\usepackage[raggedright]{titlesec}
\usepackage{epstopdf}
\definecolor{cream}{RGB}{222,217,201}

\newcommand\ba{\begin{eqnarray}}
\newcommand\ea{\end{eqnarray}}

\definecolor{Nathanblue}{rgb}{0.,0.24,0.51}
\newcommand{\blue}{\color{Nathanblue}}

\def\XXint#1#2#3{{\setbox0=\hbox{$#1{#2#3}{\int}$}
		\vcenter{\hbox{$#2#3$}}\kern-.5\wd0}}

\def\AmS{{\protect\the\textfont2
        A\kern-.1667em\lower.5ex\hbox{M}\kern-.125emS}}

\makeatletter
\def\thepage{1-\@arabic\c@page}
\def\@pnumwidth{2em}
\makeatother

\begin{document}
\title{{\blue Thermodynamic Properties of Diatomic Molecules from the Frost-Musulin Potential}}

\author{M. Parsanasab}
\affiliation{Department of Physics, College of Sciences, Yasouj University, Yasouj, 75918, Iran}

\author{R. Khordad{\,}\orcidlink{0000-0001-5180-2650}}
\email{rezakh2025@yahoo.com}
\affiliation{Department of Physics, College of Sciences, Yasouj University, Yasouj, 75918, Iran}

\author{M. Asadipour}
\affiliation{Department of Mathematics, College of Sciences, Yasouj University, Yasouj, 75918, Iran}

\author{A. Ghanbari{\,}\orcidlink{0000-0001-6147-4590}}
\affiliation{Department of Physics, College of Sciences, Yasouj University, Yasouj, 75918, Iran}

\author{V. H. Badalov{\,}\orcidlink{0000-0002-5468-1978}}
\email{badalovvatan@yahoo.com}
\affiliation{Institute for Physical Problems, Baku State University, AZ-1148 Baku, Azerbaijan}
\affiliation{Department of Theoretical Physics, Baku State University,  AZ-1148 Baku, Azerbaijan}
\date{\today}

\begin{abstract}
In this work, we examine the Frost--Musulin potential as a compact bound-region spectral model for H$_2$ and LiH and quantify how its discrete spectrum propagates into standard-state thermodynamic functions. The radial Schr\"odinger equation is treated within a near-equilibrium Pekeris representation, and the resulting finite bound-state ladder is combined with ideal-gas translational and rigid-rotor rotational factors. To clarify the model specificity, we include a direct comparison with the Morse potential matched to the same spectroscopic constants. The comparison shows that the relative performance of the Frost--Musulin and Morse spectra is molecule dependent: Morse gives a more accurate common-range $s$-state ladder for H$_2$, whereas this potential performs better for LiH. This potential thermodynamic benchmark reproduces the Gibbs free-energy deviation at sub-percent level over the studied temperature windows, while derivative-sensitive quantities such as $C_p$ and $H^\circ(T)-H^\circ_{298.15}$ show increasing high-temperature deviations. These deviations are associated with the finite bound-state ladder, the local nature of the Pekeris approximation, the factorized vibrational-plus-rigid-rotor treatment, and the absence of near-dissociative and continuum contributions. The work therefore provides a comparative benchmark and a quantitative domain of validity for FM-based bound-state thermodynamics rather than a new solution method.
\end{abstract}


\maketitle

\section{Introduction}\label{sec1}
\noindent
\lettrine[findent=2pt]{{\blue \textbf{O}}}{}btaining reliable quantum energy levels for diatomic molecules remains one of the most direct routes for connecting microscopic bonding to measurable thermal behavior{\,}\cite{1,2}. The same spectrum that governs rovibrational structure also controls the partition function and, through it, the heat capacity, entropy, free energies, and enthalpy increment{\,}\cite{12,13,14,15,18}. Since these thermodynamic observables probe the spectrum with different sensitivity, thermodynamic benchmarking provides a demanding test of whether a model potential is only locally plausible near equilibrium or predictive over a useful temperature range{\,}\cite{5,5_0,17,21}.

A broad family of molecular potentials has been developed to describe the interaction between the constituent atoms of diatomic systems{\,}\cite{19,20,21,25,26,Dong2012_MRM,Dong2007_HpIS,Wei2009_EKG,Zhang2010_RS,Miranda2010_SPT,Varshni1957,Araujo2021Review}. Their value depends not only on how accurately they reproduce the equilibrium geometry and dissociation energy, but also on whether they retain sufficient analytical transparency to connect the bound-state spectrum with thermodynamic functions{\,}\cite{19,20,25,Dong2012_MRM}. In parallel, a wide range of exact and approximate strategies has been developed for Schr\"odinger-type eigenvalue problems, underscoring the continuing importance of analytical treatments whenever the structure of the differential equation permits them{\,}\cite{7,8,8_1,9,10,11,34,Abramowitz}.

Within this landscape, the Frost--Musulin (FM) potential occupies a useful position. Originally introduced as a semiempirical representation for diatomic systems, it combines a finite dissociation energy, a well-defined equilibrium configuration, and a tunable near-equilibrium curvature within a compact analytical form{\,}\cite{27}. This balance between physical content and mathematical tractability is attractive for statistical thermodynamics, because the spectroscopic constants of a molecule can be carried transparently into the bound-state energies and then into partition-function-based observables{\,}\cite{27,29,30,31}. More broadly, the FM form belongs to the wider class of analytically useful diatomic potentials whose utility lies not only in reproducing local spectroscopy, but also in revealing how model structure propagates into thermal response{\,}\cite{Varshni1957,Araujo2021Review}.

The comparison is particularly informative for H$_2$ and LiH. These two molecules occupy markedly different regions of diatomic chemical space. H$_2$ is a compact covalent dimer with a short equilibrium bond and a large vibrational constant{\,}\cite{43}. LiH, by contrast, has a larger moment of inertia and a softer, more polar bond{\,}\cite{42}. They therefore provide a stringent two-point test of the same interaction model: one system is spectroscopically stiff and thermally compact, whereas the other samples a denser low-lying state manifold on the same temperature scale{\,}\cite{18,42,43}.

The purpose of the present work is not to introduce a new eigenvalue technique. 
The Morse oscillator already provides the classical analytical reference for 
finite anharmonic vibrational spectra of diatomic molecules{\,}\cite{Morse1929}, 
and Morse-based partition-function treatments have been used to compute thermal 
properties of H$_2$, LiH, HCl, CO, and related systems{\,}\cite{ChabiBoumali2020,17,20}. 
Analogous Schr\"odinger-equation and thermodynamic workflows have also been applied 
to Kratzer-type, Deng--Fan, Manning--Rosen, Rosen--Morse, Tietz--Hua, and other 
semiempirical diatomic potentials{\,}\cite{25,26,37,39,14,5_3,Varshni1957,Araujo2021Review}. 
The contribution of the present work is therefore comparative and diagnostic: 
we ask how the FM bound-state spectral density differs from a standard Morse 
reference and how this difference is reflected in thermodynamic observables.

In this work, we analyze the standard-state thermodynamic properties of H$_2$ and LiH within the FM potential model. After introducing a near-equilibrium Pekeris representation for the centrifugal and inverse-distance terms, the radial Schr\"odinger equation is reduced to a solvable form from which the bound-state energies are obtained analytically{\,}\cite{26,34,Abramowitz,5_2,5_3}. These energies are used to construct the vibrational partition function, which is combined with ideal-gas translational and rigid-rotor rotational contributions to obtain the total partition function and the associated thermodynamic quantities{\,}\cite{Irikura1998,Chase1998,Linstrom2001,NISTWebBookH2,NISTWebBookLiH}. The model is then assessed at two complementary levels: locally, through the FM bound-state spectrum and its comparison with the Morse reference; and globally, through comparison with NIST/JANAF Shomate reference values for $C_p^\circ(T)$, $H^\circ(T)-H^\circ_{298.15}$, and $-[G^\circ(T)-H^\circ_{298.15}]/T$.

The remainder of the paper is organized as follows. Section{\,}\ref{sec2} summarizes the theoretical framework and the solution of the radial Schr\"odinger equation under the FM potential, including the approximations employed. Section{\,}\ref{sec3} presents the spectral comparison, thermodynamic benchmark, and error analysis for H$_2$ and LiH. Section{\,}\ref{sec4} summarizes the main conclusions.

\section{FM potential, Pekeris spectrum, and thermodynamic benchmark}\label{sec2}
\noindent
The FM potential{\,}\cite{27,29,30,31} is written as
\small
\begin{equation}
    V(r)=D_e\left[
    1-
    \frac{r+\alpha r r_e-\alpha r_e^2}{r}
    e^{-\alpha(r-r_e)}
    \right],
    \label{a1}
\end{equation}
\normalsize
where $D_e$ is the dissociation energy, $\alpha$ is the screening parameter, $r_e$ is the equilibrium bond length, and $r$ is the internuclear separation. At the equilibrium point, the potential and its curvature are
\small
\begin{equation}
    V_{\mathrm{min}}=V(r_e)=0,
    \label{a3}
\end{equation}
\begin{equation}
    V''(r_e)=\alpha D_e\left(\alpha+\frac{2}{r_e}\right)>0.
    \label{a2}
\end{equation}
\normalsize
The usual constraints on a diatomic empirical potential are
\small
\begin{equation}
    \left.\frac{dV(r)}{dr}\right|_{r=r_e}=0,
    \label{a4}
\end{equation}
\begin{equation}
    V(\infty)-V(r_e)=D_e,
    \label{a5}
\end{equation}
\begin{equation}
    \left.\frac{d^2V(r)}{dr^2}\right|_{r=r_e}
    =k_e=\mu\omega_e^2,
    \label{a6}
\end{equation}
\normalsize
where $\mu$ is the reduced mass and $\omega_e$ is the angular vibrational frequency. Using the curvature condition in Eq.{\,}\eqref{a6} together with Eq.{\,}\eqref{a2}, the screening parameter is obtained from
\small
\begin{equation}
    \alpha^2+\frac{2\alpha}{r_e}
    -\frac{\mu\omega_e^2}{D_e}=0.
    \label{a7}
\end{equation}
\normalsize
The physically acceptable positive root is
\small
\begin{equation}
    \alpha=
    \frac{1}{r_e}
    \left[
    \left(1+\frac{\mu\omega_e^2r_e^2}{D_e}\right)^{1/2}
    -1
    \right].
    \label{a8}
\end{equation}
\normalsize
When the input vibrational constant is given as a wavenumber $\tilde{\nu}_e$ in cm$^{-1}$, the conversion
\[
\omega_e=2\pi c\tilde{\nu}_e
\]
is used.

\subsection{Bound states and thermodynamic model}\label{sec2_2}
\noindent
The radial Schr\"odinger equation is
\small
\begin{equation}
\frac{d^2u_{n\ell}(r)}{dr^2}
+
\frac{2\mu}{\hbar^2}
\left[
E_{n\ell}
-
V(r)
-
\frac{\hbar^2\ell(\ell+1)}{2\mu r^2}
\right]
u_{n\ell}(r)=0,
\label{a9}
\end{equation}
\normalsize
where $E_{n\ell}$ is the bound-state energy, $n$ is the vibrational quantum number, and $\ell$ is the orbital angular quantum number. In its original form, Eq.{\,}\eqref{a9} with the FM potential does not reduce directly to the hypergeometric form used below, because the potential contains an inverse-distance factor and, for $\ell\ne0$, the centrifugal term introduces an additional $r^{-2}$ contribution. We therefore use a near-equilibrium Pekeris representation for both $r^{-1}$ and $r^{-2}$ around $r=r_e${\,}\cite{Pekeris1934,Flugge1999}:
\small
\begin{equation}
\frac{1}{r^2}
\approx
\frac{1}{r_e^2}
\left(
A_0+A_1e^{-\alpha(r-r_e)}
+A_2e^{-2\alpha(r-r_e)}
\right),
\label{a10}
\end{equation}
\normalsize
and
\small
\begin{equation}
\frac{1}{r}
\approx
\frac{1}{r_e}
\left(
B_0e^{\alpha(r-r_e)}
+B_1
+B_2e^{-\alpha(r-r_e)}
\right).
\label{a11}
\end{equation}
\normalsize
The coefficients are
\small
\begin{equation}
A_0=1-\frac{3}{\alpha r_e}+\frac{3}{(\alpha r_e)^2},
\qquad
A_1=\frac{4}{\alpha r_e}-\frac{6}{(\alpha r_e)^2},
\qquad
A_2=-\frac{1}{\alpha r_e}+\frac{3}{(\alpha r_e)^2},
\label{a12}
\end{equation}
\normalsize
and
\small
\begin{equation}
B_0=-\frac{1}{2\alpha r_e}+\frac{1}{(\alpha r_e)^2},
\qquad
B_1=1-\frac{2}{(\alpha r_e)^2},
\qquad
B_2=\frac{1}{2\alpha r_e}+\frac{1}{(\alpha r_e)^2}.
\label{a13}
\end{equation}
\normalsize

Substituting Eqs.{\,}\eqref{a1}, \eqref{a10}, and \eqref{a11} into Eq.{\,}\eqref{a9} gives
\small
\begin{equation}
\frac{d^2u_{n\ell}(r)}{dr^2}
+
\alpha^2
\left(
-\epsilon
+2\beta e^{-\alpha r}
+\gamma e^{-2\alpha r}
\right)
u_{n\ell}(r)=0,
\label{a14}
\end{equation}
\normalsize
where
\small
\begin{align}
\epsilon
&=
\frac{2\mu}{\alpha^2\hbar^2}
\left[
-E_{n\ell}
+D_e(1+\alpha r_eB_0)
+\frac{\hbar^2\ell(\ell+1)}{2\mu r_e^2}A_0
\right],
\nonumber\\
\beta
&=
\frac{2\mu}{\alpha^2\hbar^2}
\left[
D_e(1+\alpha r_e-\alpha r_eB_1)
-\frac{\hbar^2\ell(\ell+1)}{2\mu r_e^2}A_1
\right]
e^{\alpha r_e},
\label{a15}\\
\gamma
&=
\frac{2\mu}{\alpha^2\hbar^2}
\left[
\alpha r_eD_eB_2
+\frac{\hbar^2\ell(\ell+1)}{2\mu r_e^2}A_2
\right]
e^{2\alpha r_e}.
\nonumber
\end{align}
\normalsize

After this approximation, the radial equation becomes analytically tractable. With the change of variable
\[
x=2\sqrt{\gamma}\,e^{-\alpha r},
\]
Eq.{\,}\eqref{a14} becomes
\small
\begin{equation}
x^2u''(x)+xu'(x)
+
\left(
-\epsilon
+\frac{\beta}{\sqrt{\gamma}}x
-\frac{x^2}{4}
\right)u(x)=0.
\label{a16}
\end{equation}
\normalsize
The physical radial function must remain regular in the short-range region and vanish in the asymptotic limit $r\rightarrow\infty$ $(x\rightarrow0)$. We therefore use the standard confluent-hypergeometric ansatz
\small
\begin{equation}
u(x)=x^{\sqrt{\epsilon}}e^{-x/2}f(x).
\label{a17}
\end{equation}
\normalsize
Substitution of Eq.{\,}\eqref{a17} into Eq.{\,}\eqref{a16} yields
\small
\begin{equation}
xf''(x)
+
\left(2\sqrt{\epsilon}+1-x\right)f'(x)
-
\left(
\sqrt{\epsilon}
+\frac{1}{2}
-\frac{\beta}{\sqrt{\gamma}}
\right)f(x)=0.
\label{a18}
\end{equation}
\normalsize
The solution can be written in terms of the confluent hypergeometric function{\,}\cite{Abramowitz},
\small
\begin{equation}
f(x)={}_1F_1(a;b;x),
\label{a19}
\end{equation}
\normalsize
with
\small
\begin{equation}
a=\sqrt{\epsilon}+\frac{1}{2}-\frac{\beta}{\sqrt{\gamma}},
\qquad
b=2\sqrt{\epsilon}+1.
\label{a20}
\end{equation}
\normalsize
The bound-state quantization follows by requiring the confluent hypergeometric series to terminate as a polynomial. This occurs for $a=-n$, with $n=0,1,2,\ldots$, namely
\small
\begin{equation}
\sqrt{\epsilon}
+\frac{1}{2}
-\frac{\beta}{\sqrt{\gamma}}
=
-n,
\qquad
n=0,1,2,\ldots .
\label{a21}
\end{equation}
\normalsize
The corresponding radial function can therefore be written as
\small
\begin{equation}
u_{n\ell}(x)
=
C_{n\ell}
x^{\sqrt{\epsilon}}
e^{-x/2}
{}_1F_1
\left(
-n;\,2\sqrt{\epsilon}+1;\,x
\right),
\label{a22}
\end{equation}
\normalsize
where $C_{n\ell}$ is the normalization constant. By inserting Eq.{\,}\eqref{a15} into the quantization condition in Eq.{\,}\eqref{a21}, the Pekeris-represented FM energy spectrum is obtained as
\small
\begin{align}
E_{n\ell}
&=
D_e(1+\alpha r_eB_0)
+
\frac{\hbar^2\ell(\ell+1)}{2\mu r_e^2}A_0
\nonumber\\
&\quad
-
\Bigg[
\frac{
D_e(1+\alpha r_e-\alpha r_eB_1)
-\dfrac{\hbar^2\ell(\ell+1)}{2\mu r_e^2}A_1
}{
2\sqrt{
\alpha r_eD_eB_2
+\dfrac{\hbar^2\ell(\ell+1)}{2\mu r_e^2}A_2
}
}
-
\frac{\alpha\hbar}{\sqrt{2\mu}}
\left(n+\frac{1}{2}\right)
\Bigg]^2 .
\label{a23}
\end{align}
\normalsize
Eq.{\,}\eqref{a23} should be understood as the energy spectrum of the Pekeris-represented FM radial problem. Its reliability is therefore controlled by the locality of the expansions in Eqs.{\,}\eqref{a10} and \eqref{a11}. The approximation is expected to be most accurate for low-lying states whose wave functions are concentrated near $r_e$, and less accurate for highly excited states that sample the dissociation-side tail of the potential.

\begin{figure*}[t]
    \centering
    \includegraphics[width=\textwidth]{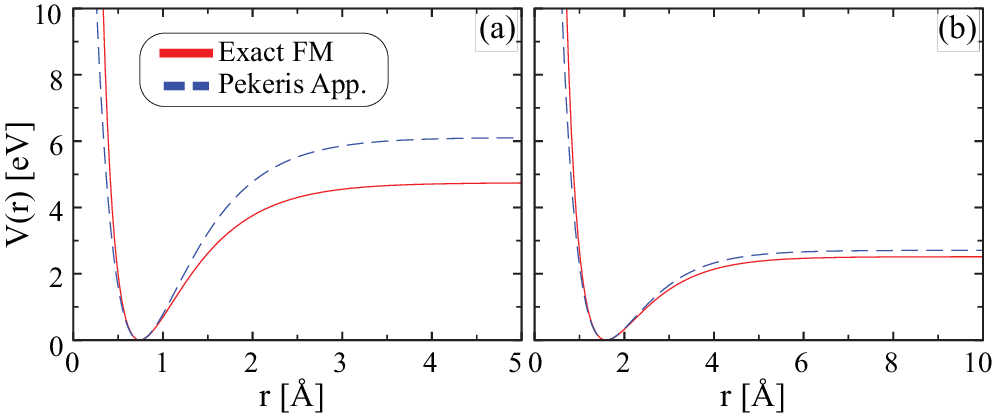}
\caption{(Color online) Exact FM potential
$V_{\mathrm{FM}}(r)$ (solid red) and its Pekeris representation
$\widetilde{V}_{\mathrm{FM}}(r)$ (dashed blue) as functions of the
internuclear separation $r$ for (a) H$_2$ and (b) LiH. The curves are
constructed from the spectroscopic parameters in Table{\,}\ref{tab:1}
and from the coefficients in Eqs.{\,}\eqref{a10}--\eqref{a13}. The
Pekeris representation reproduces the equilibrium geometry and
near-equilibrium curvature, but deviates systematically in the
large-$r$ dissociation-side tail. This emphasizes that the
approximation is local around $r_e$ and is expected to be most reliable
for low-lying bound states whose wave functions are concentrated near
the potential minimum.}
    \label{fig:1}
\end{figure*}

For the thermodynamic benchmark used in the main text, we employ the bound $s$-wave $(\ell=0)$ ladder generated by Eq.{\,}\eqref{a23}. The vibrational zero of energy is taken at the ground state, and the sum is truncated at the highest bound level satisfying $E_{n0}<D_e$. The vibrational partition function is evaluated directly as
\small
\begin{equation}
q_{\mathrm{vib}}(T)
=
\sum_{n=0}^{N_b}
\exp\!
\left[
-\beta(E_{n0}-E_{00})
\right],
\qquad
\beta=\frac{1}{k_{\mathrm B}T},
\label{a24}
\end{equation}
\normalsize
where $N_b$ denotes the largest bound $s$-wave quantum number.

For comparison with standard-state thermochemical quantities, the total ideal-gas partition function is written as
\small
\begin{align}
q_{\mathrm{trans}}(T)
&=
\left(
\frac{2\pi m k_{\mathrm B}T}{h^2}
\right)^{3/2}\cdot
\frac{k_{\mathrm B}T}{p^\circ},
\nonumber\\
q_{\mathrm{rot}}(T)
&=
\frac{1}{\sigma}
\sum_{J=0}^{\infty}
(2J+1)\cdot
\exp\!
\left[
-\frac{\hbar^2J(J+1)}{2Ik_{\mathrm B}T}
\right],
\nonumber\\
Q_{\mathrm{tot}}(T)
&=
q_{\mathrm{trans}}(T) \cdot
q_{\mathrm{rot}}(T) \cdot
q_{\mathrm{vib}}(T),
\label{a25}
\end{align}
\normalsize
where $p^\circ=1${\,}bar is the standard pressure, $m$ is the molecular mass, $\sigma$ is the rotational symmetry number, and $I=\mu r_e^2$ is the moment of inertia of the diatomic molecule.

The construction in Eq.{\,}\eqref{a25} defines the factorized vibrational-plus-rigid-rotor benchmark used in the main thermodynamic comparison. It assumes that the internal spectrum can be approximated as
\small
\begin{equation}
E_{nJ}\simeq E_{n0}+\frac{\hbar^2J(J+1)}{2I},
\label{a25b}
\end{equation}
\normalsize
and therefore neglects vibration-dependent rotational constants, centrifugal distortion, and the finite rovibrational cutoff near dissociation. A more complete bound-state treatment would replace $q_{\mathrm{rot}}q_{\mathrm{vib}}$ by the coupled rovibrational sum
\small
\begin{equation}
q_{\mathrm{rv}}(T)
=
\frac{1}{\sigma}
\sum_{n,J;\,E_{nJ}<D_e}
(2J+1)g_J
\exp\!
\left[
-\frac{E_{nJ}-E_{00}}{k_{\mathrm B}T}
\right],
\label{a25c}
\end{equation}
\normalsize
where $g_J$ denotes nuclear-spin statistical weights when required. In such a coupled formulation, $q_{\mathrm{rv}}$ replaces $q_{\mathrm{rot}}q_{\mathrm{vib}}$ and must not be multiplied again by $q_{\mathrm{rot}}$, otherwise rotational states are counted twice. This distinction is important for interpreting the $J=0+1$ diagnostic calculations reported in the Supplementary Material (SM).

The present benchmark uses the rotational symmetry number $\sigma$
but does not include separate ortho/para hydrogen thermodynamics.
Consequently, the very-low-temperature H$_2$ curves should be regarded
as illustrative, whereas the main comparison with NIST/JANAF Shomate
reference values is made over the tabulated standard-state temperature
range{\,}\cite{Chase1998,Linstrom2001,NISTWebBookH2,NISTWebBookLiH}.

Unless otherwise stated, all thermodynamic quantities reported below
are standard-state molar quantities evaluated at $p^\circ=1$ bar. For
compactness, the superscript "$\circ$" is omitted in the figure axes
and table headings. Thus, $F$, $G$, $U$, $S$, $C_V$, $C_p$, and
$H-H_{298.15}$ in figures and tables denote the corresponding
standard-state quantities
$F^\circ$, $G^\circ$, $U^\circ$, $S^\circ$, $C_V^\circ$,
$C_p^\circ$, and $H^\circ-H^\circ_{298.15}$, respectively.

The standard-state thermodynamic functions used in the following analysis are obtained from $Q_{\mathrm{tot}}$ as
\small
\begin{align}
F^\circ(T)
&=
-RT\ln Q_{\mathrm{tot}},
\label{a26a}\\
U^\circ(T)
&=
RT^2
\frac{\partial \ln Q_{\mathrm{tot}}}{\partial T},
\label{a26b}\\
S^\circ(T)
&=
R\ln Q_{\mathrm{tot}}
+
\frac{U^\circ(T)}{T},
\label{a26c}\\
H^\circ(T)
&=
U^\circ(T)+RT,
\label{a26d}\\
G^\circ(T)
&=
H^\circ(T)-TS^\circ(T),
\label{a26e}\\
C_V^\circ(T)
&=
\left(
\frac{\partial U^\circ}{\partial T}
\right)_V,
\label{a26f}\\
C_p^\circ(T)
&=
\left(
\frac{\partial H^\circ}{\partial T}
\right)_p,
\label{a26g}\\
-\frac{G^\circ(T)-H^\circ_{298.15}}{T}
&=
S^\circ(T)
-
\frac{H^\circ(T)-H^\circ_{298.15}}{T}.
\label{a26h}
\end{align}
\normalsize
Because the bound-state spectrum is analytically available, the finite sum in Eq.{\,}\eqref{a24} is evaluated directly from Eq.{\,}\eqref{a23}. Thus, Eq.{\,}\eqref{a25} defines the benchmark model used for the main thermodynamic comparison, while Eq.{\,}\eqref{a25c} identifies the physically consistent route for future improvement beyond the factorized $s$-wave plus rigid-rotor approximation.

\begin{figure*}
    \centering
    \includegraphics[width=\textwidth]{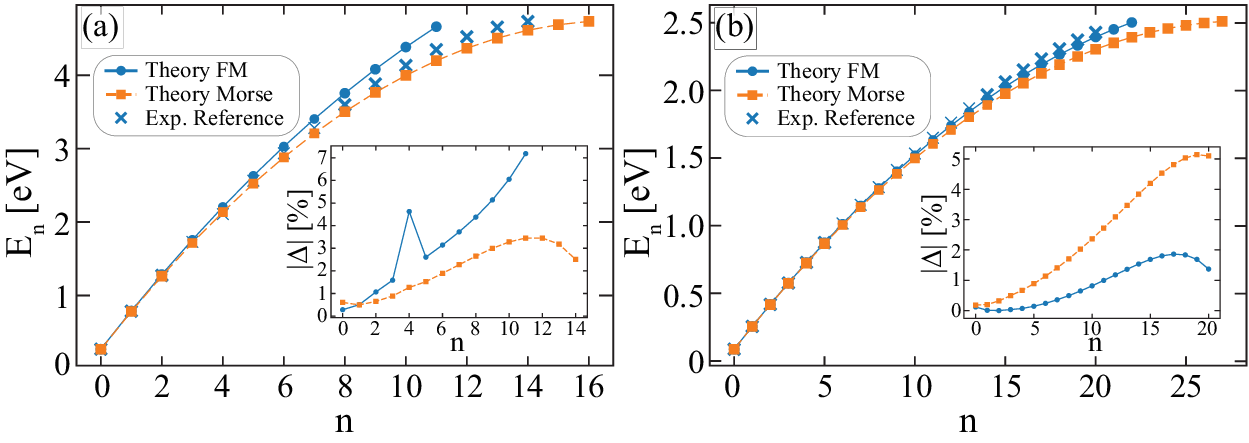}
\caption{(Color online) Spectral benchmark of the FM and Morse
$s$-state ladders against reference spectroscopic values for
(a) H$_2$ and (b) LiH. The main panels compare the absolute bound-state
energies $E_n$, while the insets show the absolute relative deviations
$|\Delta|=100|E_{\mathrm{model}}-E_{\mathrm{exp}}|/
|E_{\mathrm{exp}}|$. The comparison is molecule dependent: Morse gives
smaller common-range deviations for H$_2$, whereas FM gives smaller
common-range deviations for LiH.}
\label{fig:spectral_FM_Morse}
    \label{fig:2}
\end{figure*}

\begin{figure*}
    \centering
    \includegraphics[width=\textwidth]{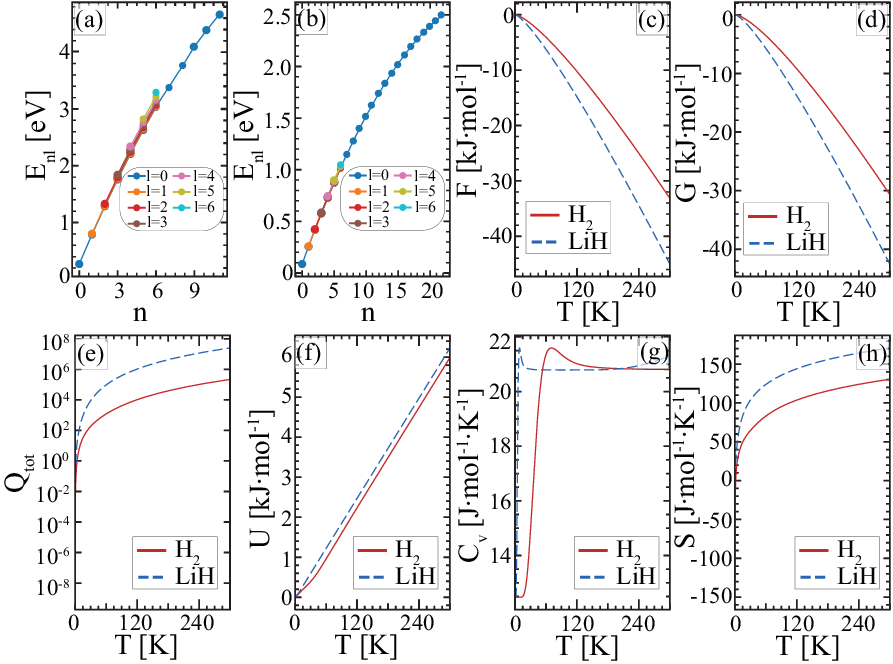}
\caption{(Color online) Bound-state spectra and illustrative
low-temperature thermodynamic trends of H$_2$ and LiH within the
FM-based factorized model. Panels (a) and (b) show the
Pekeris-represented FM eigenvalues $E_{n\ell}$ as functions of the
vibrational quantum number $n$ for different orbital quantum numbers
$\ell$. Panels (c)--(h) show the Helmholtz free energy $F$, Gibbs free
energy $G$, total partition function $Q_{\mathrm{tot}}$, internal
energy $U$, constant-volume heat capacity $C_V$, and entropy $S$ over
$0.001\leq T\leq300$ K. The thermodynamic quantities are
standard-state molar quantities; the superscript "$\circ$" is omitted
in the axis labels for compactness. Solid red curves correspond to
H$_2$, and dashed blue curves correspond to LiH. Since separate
ortho/para hydrogen thermodynamics is not included, the
very-low-temperature H$_2$ trends should be regarded as illustrative
rather than as a quantitative standard-state benchmark.}
    \label{fig:3}
\end{figure*}

\begin{figure*}
\centering
\includegraphics[width=\textwidth]{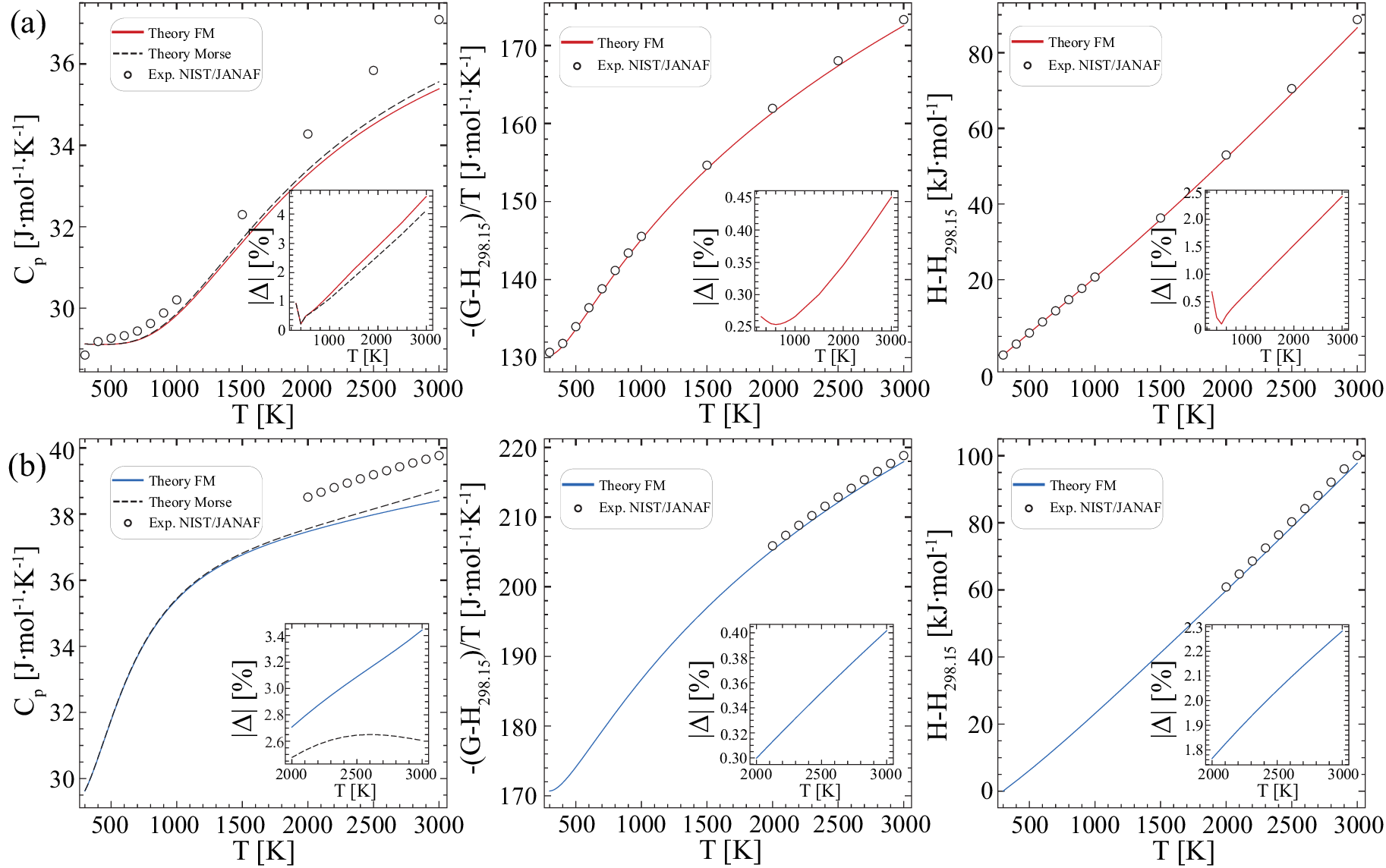}
\caption{(Color online) Temperature dependence of the standard-state
molar thermodynamic functions of (a) H$_2$ and (b) LiH compared with
NIST/JANAF Shomate reference values. The superscript "$\circ$" is
omitted in the axis labels for compactness. The first column compares
the FM and Morse heat-capacity benchmarks, while the Gibbs free-energy
deviation and enthalpy panels show the FM benchmark used for the main
thermodynamic comparison. Solid red/blue curves denote the FM model for
H$_2$/LiH, dashed black curves denote the Morse heat-capacity
benchmark, and open circles denote NIST/JANAF Shomate reference values.
Insets show absolute relative deviations in percent; in the $C_p$
panels both FM and Morse deviations are shown, while in the remaining
panels the inset corresponds to the FM benchmark.}\label{fig:4}
\end{figure*}

\begin{table}[h!]
\centering
\caption{Spectroscopic input parameters used for the FM calculations
of H$_2$ and LiH. Here $D_e$ is the dissociation energy, $r_e$ is the
equilibrium bond length, $\tilde{\nu}_e$ is the harmonic vibrational
wavenumber, $\mu$ is the reduced mass, and $\alpha$ is the FM screening
parameter obtained from the curvature-matching condition in
Eq.{\,}\eqref{a8}.}
\begin{adjustbox}{max width=\columnwidth}
\begin{tabular}{l|c|c|c|c|c|c}
\toprule \hline \hline
\textbf{Molecule} & $D_e$ & $r_e$ & $\tilde{\nu}_e$ & $\mu$ & $\alpha$ \\
                  & [eV]  & [\AA] & [cm$^{-1}$] & [a.m.u.] & [\AA$^{-1}$] \\ \hline \hline                              
\midrule
$\text{H}_2$ & 4.74460 & 0.7416    & 4395.2   & 0.50391 & 1.711487215 \\ \hline 
$\text{LiH}$ & 2.51508 & 1.5955990 & 1405.498 & 0.88013 & 1.086430866 \\ \hline \hline 
\bottomrule
\end{tabular}
\end{adjustbox}
\label{tab:1}
\end{table}

\begin{table*}[t]
\centering
\scriptsize
\caption{Low-lying FM and Morse bound-state energies of
H$_2$ and LiH. The FM values are obtained from
Eq.{\,}\eqref{a23}, while the Morse values are computed using the same
molecular input constants. For $\ell=0$, reference spectroscopic
energies are included and absolute relative deviations are reported.
The H$_2$ reference energies are taken from Dabrowski{\,}\cite{Dabrowski1984_LymanWernerH2},
and the LiH reference energies are taken from Coxon and
Dickinson{\,}\cite{CoxonDickinson2004_LiH_DPF}. The comparison shows
that the relative performance of the FM and Morse spectra
is molecule dependent.}
\label{tab:2}
\begin{adjustbox}{max width=\textwidth}
\begin{tabular}{cc|ccccc|ccccc}
\toprule\toprule
& & \multicolumn{5}{c|}{H$_2$} & \multicolumn{5}{c}{LiH}\\
\cmidrule(lr){3-7}\cmidrule(lr){8-12}
$n$ & $\ell$ & FM & Morse & Exp. & $|\Delta_{\rm FM}|$ & $|\Delta_{\rm M}|$ & FM & Morse & Exp. & $|\Delta_{\rm FM}|$ & $|\Delta_{\rm M}|$\\
& & (eV) & (eV) & (eV) & (\%) & (\%) & (eV) & (eV) & (eV) & (\%) & (\%)\\
\midrule
0 & 0 & 0.269431970 & 0.268557567 & 0.2701925 & 0.28 & 0.61 & 0.086429558 & 0.086375695 & 0.08653316 & 0.12 & 0.18 \\
1 & 0 & 0.790071476 & 0.782201953 & 0.7860960 & 0.51 & 0.50 & 0.255084149 & 0.254599378 & 0.25511164 & 0.01 & 0.20 \\
1 & 1 & 0.804627593 & 0.787680432 & -- & -- & -- & 0.256883350 & 0.255083248 & -- & -- & -- \\
2 & 0 & 1.286411757 & 1.264552003 & 1.2728210 & 1.07 & 0.65 & 0.418132709 & 0.416786122 & 0.41812872 & 0.00 & 0.32 \\
2 & 1 & 1.300631570 & 1.270030482 & -- & -- & -- & 0.419887971 & 0.417269992 & -- & -- & -- \\
2 & 2 & 1.328947747 & 1.280987435 & -- & -- & -- & 0.423396051 & 0.418237735 & -- & -- & -- \\
3 & 0 & 1.758452807 & 1.715607718 & 1.7309850 & 1.59 & 0.89 & 0.575575239 & 0.572935931 & 0.57576978 & 0.03 & 0.49 \\
3 & 1 & 1.772336317 & 1.721086197 & -- & -- & -- & 0.577286560 & 0.573419801 & -- & -- & -- \\
3 & 2 & 1.799985393 & 1.732043149 & -- & -- & -- & 0.580706812 & 0.574387544 & -- & -- & -- \\
3 & 3 & 1.841168847 & 1.748478580 & -- & -- & -- & 0.585831228 & 0.575839154 & -- & -- & -- \\
4 & 0 & 2.206194629 & 2.135369097 & 2.1086590 & 4.63 & 1.27 & 0.727411737 & 0.723048800 & 0.72792287 & 0.07 & 0.67 \\
4 & 1 & 2.219741836 & 2.140847575 & -- & -- & -- & 0.729079119 & 0.723532670 & -- & -- & -- \\
4 & 2 & 2.246723812 & 2.151804528 & -- & -- & -- & 0.732411542 & 0.724500412 & -- & -- & -- \\
4 & 3 & 2.286920190 & 2.168239958 & -- & -- & -- & 0.737404336 & 0.725952022 & -- & -- & -- \\
4 & 4 & 2.340011395 & 2.190153863 & -- & -- & -- & 0.744050514 & 0.727887505 & -- & -- & -- \\
5 & 0 & 2.629637222 & 2.523836140 & 2.5628460 & 2.61 & 1.52 & 0.873642204 & 0.867124730 & 0.87489678 & 0.14 & 0.89 \\
5 & 1 & 2.642848126 & 2.529314618 & -- & -- & -- & 0.875265648 & 0.867608600 & -- & -- & -- \\
5 & 2 & 2.669163002 & 2.540271570 & -- & -- & -- & 0.878510241 & 0.868576343 & -- & -- & -- \\
5 & 3 & 2.708372305 & 2.556707000 & -- & -- & -- & 0.883371414 & 0.870027953 & -- & -- & -- \\
5 & 4 & 2.760172230 & 2.578620904 & -- & -- & -- & 0.889842326 & 0.871963435 & -- & -- & -- \\
5 & 5 & 2.824176446 & 2.606013286 & -- & -- & -- & 0.897913920 & 0.874382787 & -- & -- & -- \\
6 & 0 & 3.028780587 & 2.881008848 & 2.9365010 & 3.14 & 1.89 & 1.014266641 & 1.005163723 & 1.01669967 & 0.24 & 1.13 \\
6 & 1 & 3.041655187 & 2.886487326 & -- & -- & -- & 1.015846145 & 1.005647593 & -- & -- & -- \\
6 & 2 & 3.067302962 & 2.897444278 & -- & -- & -- & 1.019002908 & 1.006615335 & -- & -- & -- \\
6 & 3 & 3.105525190 & 2.913879707 & -- & -- & -- & 1.023732461 & 1.008066945 & -- & -- & -- \\
6 & 4 & 3.156033838 & 2.935793610 & -- & -- & -- & 1.030028107 & 1.010002428 & -- & -- & -- \\
6 & 5 & 3.218462770 & 2.963185992 & -- & -- & -- & 1.037880985 & 1.012421779 & -- & -- & -- \\
6 & 6 & 3.292381425 & 2.996056848 & -- & -- & -- & 1.047280072 & 1.015325001 & -- & -- & -- \\

\midrule
\multicolumn{2}{c|}{$s$-state MAPE, $n=0$--6} & \multicolumn{2}{c}{FM: 1.97\%, Morse: 1.05\%} & \multicolumn{3}{c|}{} & \multicolumn{2}{c}{FM: 0.09\%, Morse: 0.56\%} & \multicolumn{3}{c}{}\\
\bottomrule\bottomrule
\end{tabular}
\end{adjustbox}
\label{tab:2}
\end{table*}

\begin{table}[t]
\squeezetable
\caption{Standard-state molar thermodynamic benchmark for H$_2$.
For compactness, the superscript "$\circ$" is omitted in the table
headings. Reference experimental values are evaluated from the NIST/JANAF Shomate
polynomial expressions for $C_p(T)$,
$H(T)-H_{298.15}$, and $S(T)$, from which
$-[G(T)-H_{298.15}]/T$ is obtained. The theoretical values are computed
using the factorized FM model
$Q_{\mathrm{tot}}=q_{\mathrm{trans}}q_{\mathrm{rot}}q_{\mathrm{vib}}$,
where $q_{\mathrm{vib}}$ is constructed from the finite FM $s$-wave
bound-state ladder. Percentages listed beneath each row are absolute
relative deviations from the NIST/JANAF reference values.}\label{tab:3}
\begin{adjustbox}{max width=\columnwidth}
\footnotesize
\begin{ruledtabular}
\begin{tabular}{c|cc|cc|cc}
& \multicolumn{2}{c|}{$C_p$} & \multicolumn{2}{c|}{$-\,[G-H_{298.15}]/T$} & \multicolumn{2}{c}{$H-H_{298.15}$} \\
Temp. & \multicolumn{2}{c|}{[J mol$^{-1}$ K$^{-1}$]} & \multicolumn{2}{c|}{[J mol$^{-1}$ K$^{-1}$]} & \multicolumn{2}{c}{[kJ mol$^{-1}$]} \\
$T$ (K) & Exp. & Theor. & Exp. & Theor. & Exp. & Theor. \\
\hline
300 & 28.8495 & 29.1208 & 130.6802 & 130.3326 & 0.053510 & 0.053874 \\
& & 0.94\% & & 0.27\% & & 0.68\% \\
\hline
400 & 29.1816 & 29.1118 & 131.8175 & 131.4747 & 2.959150 & 2.965413 \\
& & 0.24\% & & 0.26\% & & 0.21\% \\
\hline
500 & 29.2618 & 29.1140 & 133.9732 & 133.6311 & 5.882022 & 5.876578 \\
& & 0.51\% & & 0.26\% & & 0.09\% \\
\hline
600 & 29.3246 & 29.1407 & 136.3922 & 136.0459 & 8.811080 & 8.789020 \\
& & 0.63\% & & 0.25\% & & 0.25\% \\
\hline
700 & 29.4392 & 29.2144 & 138.8216 & 138.4678 & 11.748712 & 11.706299 \\
& & 0.76\% & & 0.25\% & & 0.36\% \\
\hline
800 & 29.6250 & 29.3524 & 141.1712 & 140.8078 & 14.701308 & 14.634066 \\
& & 0.92\% & & 0.26\% & & 0.46\% \\
\hline
900 & 29.8825 & 29.5595 & 143.4113 & 143.0366 & 17.676109 & 17.579102 \\
& & 1.08\% & & 0.26\% & & 0.55\% \\
\hline
1000 & 30.2041 & 29.8290 & 145.5360 & 145.1487 & 20.679951 & 20.548057 \\
& & 1.24\% & & 0.27\% & & 0.64\% \\
\hline
1500 & 32.2990 & 31.6210 & 154.6523 & 154.1875 & 36.290283 & 35.896166 \\
& & 2.10\% & & 0.30\% & & 1.09\% \\
\hline
2000 & 34.2791 & 33.2868 & 161.9426 & 161.3829 & 52.951303 & 52.140696 \\
& & 2.89\% & & 0.35\% & & 1.53\% \\
\hline
2500 & 35.8405 & 34.5136 & 168.0435 & 167.3772 & 70.498107 & 69.108079 \\
& & 3.70\% & & 0.40\% & & 1.97\% \\
\hline
3000 & 37.0890 & 35.3886 & 173.3111 & 172.5296 & 88.741400 & 86.595618 \\
& & 4.58\% & & 0.45\% & & 2.42\% \\
\hline
3500 & 38.1452 & 36.0284 & 177.9598 & 177.0553 & 107.554764 & 104.457653 \\
& & 5.55\% & & 0.51\% & & 2.88\% \\
\hline
4000 & 39.1167 & 36.5155 & 182.1292 & 181.0948 & 126.872982 & 122.598720 \\
& & 6.65\% & & 0.57\% & & 3.37\% \\
\hline
4500 & 40.0195 & 36.9006 & 185.9164 & 184.7448 & 146.660214 & 140.956257 \\
& & 7.79\% & & 0.63\% & & 3.89\% \\
\hline
5000 & 40.8280 & 37.2122 & 189.3915 & 188.0756 & 166.876885 & 159.487142 \\
& & 8.86\% & & 0.69\% & & 4.43\% \\
\hline
5500 & 41.4960 & 37.4648 & 192.6064 & 191.1395 & 187.464869 & 178.158693 \\
& & 9.71\% & & 0.76\% & & 4.96\% \\
\hline
6000 & 41.9669 & 37.6643 & 195.6000 & 193.9769 & 208.340082 & 196.943121 \\
& & 10.25\% & & 0.83\% & & 5.47\% \\
\hline
\end{tabular}
\end{ruledtabular}
\end{adjustbox}
\end{table}

\begin{table}[t]
\squeezetable
\caption{Standard-state thermodynamic benchmark for LiH over the
NIST/JANAF gas-phase Shomate temperature range. For compactness, the superscript "$\circ$" is omitted in the table
headings. Reference experimental values are evaluated from the NIST/JANAF Shomate
polynomial expressions for $C_p(T)$,
$H(T)-H_{298.15}$, and $S(T)$, from which
$-[G(T)-H_{298.15}]/T$ is obtained. The theoretical values are computed
using the factorized FM model
$Q_{\mathrm{tot}}=q_{\mathrm{trans}}q_{\mathrm{rot}}q_{\mathrm{vib}}$,
where $q_{\mathrm{vib}}$ is constructed from the finite FM $s$-wave
bound-state ladder. Percentages listed beneath each row are absolute
relative deviations from the NIST/JANAF reference values.}
\label{tab:4}
\footnotesize
\begin{ruledtabular}
\begin{tabular}{c|cc|cc|cc}
& \multicolumn{2}{c|}{$C_p$} & \multicolumn{2}{c|}{$-\,[G-H_{298.15}]/T$} & \multicolumn{2}{c}{$H-H_{298.15}$} \\
Temp. & \multicolumn{2}{c|}{[J mol$^{-1}$ K$^{-1}$]} & \multicolumn{2}{c|}{[J mol$^{-1}$ K$^{-1}$]} & \multicolumn{2}{c}{[kJ mol$^{-1}$]} \\
$T$ (K) & Exp. & Theor. & Exp. & Theor. & Exp. & Theor. \\
\hline
2000 & 38.5133 & 37.4707 & 205.8766 & 205.2596 & 60.874095 & 59.799515 \\
& & 2.71\% & & 0.30\% & & 1.77\% \\
\hline
2100 & 38.6611 & 37.5819 & 207.3712 & 206.7273 & 64.732880 & 63.552197 \\
& & 2.79\% & & 0.31\% & & 1.82\% \\
\hline
2200 & 38.8016 & 37.6876 & 208.8135 & 208.1430 & 68.606070 & 67.315714 \\
& & 2.87\% & & 0.32\% & & 1.88\% \\
\hline
2300 & 38.9361 & 37.7888 & 210.2072 & 209.5100 & 72.493003 & 71.089569 \\
& & 2.95\% & & 0.33\% & & 1.94\% \\
\hline
2400 & 39.0654 & 37.8862 & 211.5553 & 210.8317 & 76.393120 & 74.873350 \\
& & 3.02\% & & 0.34\% & & 1.99\% \\
\hline
2500 & 39.1904 & 37.9802 & 212.8607 & 212.1107 & 80.305946 & 78.666698 \\
& & 3.09\% & & 0.35\% & & 2.04\% \\
\hline
2600 & 39.3116 & 38.0710 & 214.1259 & 213.3498 & 84.231076 & 82.469287 \\
& & 3.16\% & & 0.36\% & & 2.09\% \\
\hline
2700 & 39.4297 & 38.1587 & 215.3535 & 214.5514 & 88.168170 & 86.280799 \\
& & 3.22\% & & 0.37\% & & 2.14\% \\
\hline
2800 & 39.5452 & 38.2431 & 216.5455 & 215.7176 & 92.116936 & 90.100915 \\
& & 3.29\% & & 0.38\% & & 2.19\% \\
\hline
2900 & 39.6584 & 38.3239 & 217.7040 & 216.8505 & 96.077133 & 93.929295 \\
& & 3.36\% & & 0.39\% & & 2.24\% \\
\hline
3000 & 39.7698 & 38.4010 & 218.8309 & 217.9520 & 100.048561 & 97.765573 \\
& & 3.44\% & & 0.40\% & & 2.28\% \\
\hline
3500 & 40.3113 & 38.7159 & 224.0463 & 223.0424 & 120.069567 & 117.050146 \\
& & 3.96\% & & 0.45\% & & 2.51\% \\
\hline
4000 & 40.8513 & 38.8829 & 228.6799 & 227.5537 & 140.359636 & 136.456698 \\
& & 4.82\% & & 0.49\% & & 2.78\% \\
\hline
4500 & 41.4178 & 38.8808 & 232.8525 & 231.6037 & 160.925260 & 155.904692 \\
& & 6.13\% & & 0.54\% & & 3.12\% \\
\hline
5000 & 42.0350 & 38.7180 & 236.6517 & 235.2764 & 181.785854 & 175.310598 \\
& & 7.89\% & & 0.58\% & & 3.56\% \\
\hline
5500 & 42.7254 & 38.4230 & 240.1429 & 238.6339 & 202.972448 & 194.600629 \\
& & 10.07\% & & 0.63\% & & 4.12\% \\
\hline
6000 & 43.5105 & 38.0317 & 243.3763 & 241.7233 & 224.527025 & 213.717553 \\
& & 12.59\% & & 0.68\% & & 4.81\% \\
\hline
\end{tabular}
\end{ruledtabular}
\end{table}

\begin{table}[t]
\caption{Error hierarchy of the FM thermodynamic benchmark over
moderate and extended temperature windows. For compactness, the superscript "$\circ$" is omitted in the table
headings. The two rows for each
molecule differ only by the temperature interval used in the averaging;
the potential, spectrum, and factorized partition-function model are
unchanged. Entries are mean/max absolute relative deviations in
percent.}
\label{tab:5}
\centering
\begin{tabular}{lccc}
\hline\hline
System and range & $C_p$ &
$-[G-H_{298.15}]/T$ &
$H-H_{298.15}$\\
\hline
H$_2$, 300--3000 K & 1.63 / 4.58 & 0.30 / 0.45 & 0.85 / 2.42\\
H$_2$, 300--6000 K & 3.80 / 10.25 & 0.42 / 0.83 & 1.96 / 5.47\\
LiH, 2000--3000 K & 3.08 / 3.44 & 0.35 / 0.40 & 2.03 / 2.28\\
LiH, 2000--6000 K & 4.67 / 12.59 & 0.43 / 0.68 & 2.55 / 4.81\\
\hline\hline
\end{tabular}
\end{table}

\section{Results and Discussion}\label{sec3}
\noindent
The spectroscopic constants listed in Table{\,}\ref{tab:1} place H$_2$ and LiH on markedly different molecular energy scales. H$_2$ combines a short equilibrium bond length with a large harmonic vibrational wavenumber $\tilde{\nu}_e$, whereas LiH is characterized by a substantially longer bond and a much smaller $\tilde{\nu}_e$. Through Eq.{\,}\eqref{a8}, these quantities determine the screening parameter $\alpha$ and, with it, the local curvature of the FM well and the Pekeris coefficients entering the effective radial problem. Table{\,}\ref{tab:1} therefore plays a structural role in the analysis: it fixes the hierarchy of vibrational and rotational energy scales from which both the bound-state spectrum and the subsequent thermodynamic response emerge{\,}\cite{27, Varshni1957}.

The first test of the analytical construction is provided by Fig.{\,}\ref{fig:1}, which compares the exact FM potential with its Pekeris representation for H$_2$ and LiH. In both systems, the two curves are nearly indistinguishable in the vicinity of the minimum and throughout the portion of configuration space sampled by the low-lying bound states. The approximation is therefore local rather than global. It preserves the equilibrium geometry and the near-equilibrium curvature that govern the low-lying discrete spectrum, while gradually departing from the exact dissociation-side tail at larger separations. This is precisely the regime for which the Pekeris expansion is intended: it regularizes the inverse-distance and centrifugal terms in a controlled neighborhood of $r=r_e$, where the low-lying bound-state wave functions carry most of their weight, rather than attempting to reproduce the full long-range form of the potential{\,}\cite{27, Varshni1957}.

To clarify the model specificity of the FM spectrum, Table{\,}\ref{tab:2} and Fig.{\,}\ref{fig:2} compare the low-lying FM bound-state energies with a Morse reference constructed from the same molecular input constants. For the $s$-state ladder, available reference spectroscopic energies are also included, together with absolute relative deviations. The comparison is not intended to establish universal superiority of either potential. Rather, it identifies how the spectral-density error depends on the molecule and on the shape of the potential.

The FM--Morse comparison in Table{\,}\ref{tab:2} and Fig.{\,}\ref{fig:2} shows that the relative spectral performance is molecule dependent. For H$_2$, the Morse ladder gives smaller common-range $s$-state deviations than the FM ladder. Over the common range $n=0$--11, the mean absolute percentage error is 1.83\% for Morse and 3.36\% for FM. For LiH, the trend is reversed: over the common range $n=0$--20, the FM ladder gives a mean absolute percentage error of 0.87\%, whereas Morse gives 2.54\%. Thus, the FM potential is not presented as uniformly superior to the Morse reference. Its advantage is molecule dependent and is most visible here in the LiH $s$-state spectral ladder, whereas Morse remains the more accurate common-range reference for H$_2$.

The orbital-state-resolved FM spectra and the illustrative low-temperature thermodynamic trends are summarized in Fig.{\,}\ref{fig:3}. Panels (a) and (b) show the FM eigenvalues $E_{n\ell}$ as functions of the vibrational quantum number $n$ for different orbital quantum numbers $\ell$. For both molecules, the eigenvalues increase with $n$ and $\ell$ over the range considered. H$_2$ exhibits wider vibrational spacings because its larger $\tilde{\nu}_e$ and smaller reduced mass generate a stiffer effective well. LiH, by contrast, supports a denser low-lying ladder because its larger moment of inertia and softer vibrational scale lower the excitation thresholds. The $\ell$-dependent splitting follows the same logic: since the relevant rotational scale varies inversely with $\mu r_e^2$, the splitting is more pronounced in H$_2$ than in LiH at comparable excitation.

The low-temperature trends in Fig.{\,}\ref{fig:3}(c)--(h) are obtained from
\begin{equation}
Q_{\mathrm{tot}}(T)=q_{\mathrm{trans}}(T)\cdot q_{\mathrm{rot}}(T)\cdot q_{\mathrm{vib}}(T).
\end{equation}
Here the vibrational contribution is generated from the FM $s$-wave bound-state spectrum, while the translational and rotational factors follow the standard ideal-gas and rigid-rotor constructions{\,}\cite{Irikura1998, Chase1998, Linstrom2001}. These low-temperature curves are useful for visualizing the internal consistency of the factorized model. Since separate ortho/para hydrogen thermodynamics is not included, the very-low-temperature H$_2$ behavior should be regarded as illustrative rather than as a quantitative standard-state benchmark.

Because the standard-state formulation includes translational and rotational reference-state terms in addition to the internal vibrational spectrum, the absolute low-temperature magnitudes of $F^\circ$, $G^\circ$, and $S^\circ$ should be read as standard-state thermodynamic quantities rather than as purely spectroscopic observables. Within that convention, however, the trends in Fig.{\,}\ref{fig:3}(c)--(h) are internally consistent and chemically meaningful.

A more demanding assessment of the framework is furnished by the comparison with NIST/JANAF Shomate reference values in Fig.{\,}\ref{fig:4} and Tables{\,}\ref{tab:3}--\ref{tab:4}. The figure contains two levels of comparison. The full thermodynamic benchmark is shown for the FM model, while the first column additionally includes a Morse heat-capacity benchmark constructed with the same translational and rigid-rotor factors. Therefore, the FM--Morse comparison in $C_p^\circ$ isolates the effect of replacing the FM vibrational ladder by the Morse ladder while keeping the remaining statistical-mechanical framework unchanged.

Across the full overlap window, the theoretical curves follow the NIST/JANAF trends smoothly and with systematic, chemically interpretable bias rather than irregular point-to-point scatter. This behavior shows that the residual discrepancy is controlled by the physical scope of the model rather than by numerical instability. In practical terms, the benchmark probes how far a finite bound-state vibrational description, combined with standard translational and rigid-rotor ideal-gas contributions, can reproduce the standard thermochemical response of real diatomic molecules{\,}\cite{Irikura1998, Chase1998, Linstrom2001, NISTWebBookH2, NISTWebBookLiH}.

For H$_2$, over the 300--3000{\,}K interval, the FM benchmark gives mean/max deviations of 1.63\%/4.58\% for $C_p^\circ$, 0.30\%/0.45\% for $-[G^\circ-H^\circ_{298.15}]/T$, and 0.85\%/2.42\% for $H^\circ-H^\circ_{298.15}$, see Fig.{\,}\ref{fig:4} and and Tables{\,}\ref{tab:3}--\ref{tab:5}. In the same $C_p^\circ$ window, the Morse benchmark gives a slightly smaller mean/max deviation of 1.48\%/4.11\%. Thus, Morse improves the heat-capacity agreement modestly, but the difference is not qualitative. Both models reproduce the overall thermal activation trend, while the remaining discrepancy is amplified in $C_p^\circ$ because this quantity is sensitive to the curvature of the thermal population and to the upper part of the bound ladder. Over the extended 300--6000{\,}K interval, the same trend persists: the FM $C_p^\circ$ deviation increases to 3.80\%/10.25\%, while the Morse value is 3.31\%/8.33\%.

LiH displays the same hierarchy of observables, but in a more demanding spectral regime. Over 2000--3000{\,}K, the FM benchmark gives mean/max deviations of 3.08\%/3.44\% for $C_p^\circ$, 0.35\%/0.40\% for $-[G^\circ-H^\circ_{298.15}]/T$, and 2.03\%/2.28\% for $H^\circ-H^\circ_{298.15}$. The Morse heat-capacity benchmark again gives a somewhat smaller $C_p^\circ$ deviation, 2.60\%/2.65\%, over the same temperature window. Across the full 2000--6000{\,}K interval, the FM $C_p^\circ$ deviation reaches 4.67\%/12.59\%, while the Morse value is 3.44\%/9.22\%. This confirms that $C_p^\circ$ is the most sensitive observable in the present comparison.

This result should not be interpreted as a simple ranking of the two potentials at all levels. At the spectral level, the FM $s$-state ladder is more accurate for LiH over the available common range, as shown in Table{\,}\ref{tab:2} and Fig.{\,}\ref{fig:2}. At the thermodynamic level, however, $C_p^\circ$ depends on the distribution of level spacings and on the thermal accessibility of the upper bound states. Therefore, a potential can be competitive or even better at the level-energy benchmark while still giving a slightly larger heat-capacity deviation. This observable dependence is one of the main lessons of the FM--Morse comparison.

Taken together, Fig.{\,}\ref{fig:4} and Tables{\,}\ref{tab:3}--\ref{tab:5} show an observable-dependent accuracy hierarchy. The Gibbs free-energy deviation is the most stable quantity and remains at sub-percent level over all tested windows. By contrast, $C_p^\circ$ and $H^\circ-H^\circ_{298.15}$ show increasing high-temperature deviations, with the largest errors concentrated in $C_p^\circ$. The Morse heat-capacity comparison in Fig.{\,}\ref{fig:4} confirms that this sensitivity is controlled by the thermal spectral density rather than only by the near-equilibrium potential shape. The FM model therefore should not be interpreted as a globally superior thermochemical model. Its strength lies in providing a compact bound-region spectral representation whose domain of validity can be quantified against both Morse and NIST/JANAF reference values.

A supplementary diagnostic comparison is provided in Fig.{\,}1S of the SM to illustrate why adding only the first explicitly resolved rotational channel does not constitute a complete improvement of the thermodynamic model. In that diagnostic, the benchmark factorized model based on the FM $s$-wave vibrational ladder and the full rigid-rotor partition function is compared with a truncated explicit $J=0+1$ bound-state sum. The $J=0+1$ truncation moves the calculation in the expected direction but still omits the thermally active higher-$J$ manifold. This confirms that the next physical extension should not be a two-channel truncation, but a full coupled rovibrational sum over all bound levels satisfying $E_{nJ}<D_e$.

From a broader perspective, the present results identify the next physically meaningful route for improvement. The main limitation is not removed by further algebraic manipulation of the same $s$-wave ladder. A more complete treatment should replace the factorized $q_{\mathrm{rot}}q_{\mathrm{vib}}$ internal partition function by the coupled bound rovibrational sum introduced earlier, including vibration-dependent rotational constants, centrifugal distortion, and the finite rovibrational cutoff near dissociation. At still higher temperatures, near-dissociative and continuum contributions will also be required. This identifies the FM framework as a transparent bound-region benchmark rather than a complete high-temperature thermochemical model.

\section{Conclusions}\label{sec4}
\noindent
We have assessed the FM potential as a compact bound-region spectral model for H$_2$ and LiH and traced how its finite bound-state ladder propagates into standard-state thermodynamic observables. The Pekeris-represented FM spectrum was combined with ideal-gas translational and rigid-rotor rotational factors and benchmarked against both a Morse reference and NIST/JANAF Shomate thermochemistry. The comparison shows that the spectral performance is molecule dependent: Morse gives the smaller common-range $s$-state error for H$_2$, whereas the FM ladder gives the smaller error for LiH.

The thermodynamic benchmark reveals an observable-dependent accuracy hierarchy. The Gibbs free-energy deviation remains at sub-percent level over the tested temperature windows, whereas $C_p$ and $H-H_{298.15}$ show increasing deviations at elevated temperature. The Morse heat-capacity comparison gives slightly smaller $C_p$ deviations, indicating that heat capacity is governed not only by individual low-lying level errors, but also by the level-spacing distribution and the density of upper bound states. The residual high-temperature errors originate from the finite bound-state ladder, the local nature of the Pekeris approximation, the factorized vibrational-plus-rigid-rotor treatment, and the absence of
non-rigid rotational, near-dissociative, and continuum contributions. The physically consistent extension is therefore a full coupled bound rovibrational partition function over all states satisfying $E_{nJ}<D_e$, followed at higher temperature by dissociative and continuum corrections. Thus, the FM potential provides a transparent bound-region benchmark and a clear starting point for more complete high-temperature molecular thermodynamics, rather than a new solution method or a universally complete thermochemical model.

\section*{Acknowledgments}
\noindent

\section*{Data availability statement}
\noindent All data that support the findings of this study are included within the article and supplementary material.

\section*{Conflict of interest}
\noindent The authors declare no competing interests. All research has been carried out within an appropriate ethical framework.

\bibliographystyle{apsrev4-2}
\bibliography{references}

\clearpage
\newpage

\setcounter{equation}{0}
\setcounter{figure}{0}
\setcounter{table}{0}
\setcounter{page}{1}
\renewcommand{\theequation}{S\arabic{equation}}
\renewcommand{\thefigure}{S\arabic{figure}}
\renewcommand{\thetable}{S\arabic{table}}

\end{document}